\documentclass[preprint, prb,superscriptaddress, showpacs,11pt]{revtex4-1}
\usepackage[utf8]{inputenc}
\usepackage{amsmath}
\usepackage{amssymb}
\usepackage{multirow}
\usepackage{graphicx}
\usepackage{natbib}
\usepackage{color}

\begin{document}
\title{First-principle study of paraelectric and ferroelectric CsH$_2$PO$_4$ including dispersion forces: stability and related vibrational, dielectric and elastic properties}

\author{Benoit Van Troeye}
\email{benoit.vantroeye@uclouvain.be}
\affiliation{Institute for Condensed Matter and Nanosciences, European Theoretical Spectroscopy Facility, Universit\'{e} catholique de Louvain, Chemin des étoiles 8, B-1348 Louvain-la-Neuve, Belgium}

\author{Michiel Jan van Setten}
\affiliation{Institute for Condensed Matter and Nanosciences, European Theoretical Spectroscopy Facility, Universit\'{e} catholique de Louvain, Chemin des étoiles 8, B-1348 Louvain-la-Neuve, Belgium}

\author{Matteo Giantomassi}
\affiliation{Institute for Condensed Matter and Nanosciences, European Theoretical Spectroscopy Facility, Universit\'{e} catholique de Louvain, Chemin des étoiles 8, B-1348 Louvain-la-Neuve, Belgium}

\author{Marc Torrent}
\affiliation{CEA, DAM, DIF, F-91297 Arpajon, France}

\author{Gian-Marco Rignanese}
\affiliation{Institute for Condensed Matter and Nanosciences, European Theoretical Spectroscopy Facility, Universit\'{e} catholique de Louvain, Chemin des étoiles 8, B-1348 Louvain-la-Neuve, Belgium}

\author{Xavier Gonze}
\affiliation{Institute for Condensed Matter and Nanosciences, European Theoretical Spectroscopy Facility, Universit\'{e} catholique de Louvain, Chemin des étoiles 8, B-1348 Louvain-la-Neuve, Belgium}

\begin{abstract}
Using density functional theory (DFT) and density functional perturbation theory (DFPT), we investigate the stability and response functions of CsH$_2$PO$_4$, a ferroelectric material at low temperature.
This material cannot be described properly by the usual (semi-)local approximations within DFT. The long-range e$^-$-e$^-$ correlation needs to be properly taken into account, using, for instance,
Grimme's DFT-D methods, as investigated in this work. We find that DFT-D3(BJ) performs the best for the members of the dihydrogenated alkali  phosphate family (KH$_2$PO$_4$, RbH$_2$PO$_4$, CsH$_2$PO$_4$), leading to experimental lattice parameters reproduced with
an average deviation of 0.5 \%. With these DFT-D methods, the structural, dielectric, vibrational and mechanical properties of CsH$_2$PO$_4$ are globally in excellent agreement with the available experiments ($<$ 2\% MAPE for Raman-active phonons).
Our study suggests the possible existence of a new low-temperature phase for CsH$_2$PO$_4$, not yet reported experimentally. Finally, we report the implementation of DFT-D
contributions to elastic constants within DFPT.
\end{abstract}

\maketitle

\section*{Introduction}


 In recent years, CsH$_2$PO$_4$ (CDP), a ferroelectric material, has received a renewed interest because of its possible use as solid electrolyte 
 in fuel cells \cite{Haile2007,Qing2015,Qing2014}. Indeed, its high-temperature cubic form has been shown to behave as a superprotonic phase \cite{Kim2015}. 
 
 The CDP crystal, and its deuterated counterpart CsD$_2$PO$_4$ (DCDP), crystallize at low temperature into a $P2_1$ ferroelectric phase. 
 CDP and DCDP undergo a phase transition, accompanied by an increase of symmetry, from the ferroelectric phase
 to a $P2_1/m$ paraelectric one at around 154.5K and 264K respectively \cite{Levstik1975,Matsunga1980,Itoh1983,Frazer1979}. Because of these phase transitions and associated ferroelectric and superprotonic properties, 
 the structural, electric and dynamical properties of this material are quite complex. Complementing the available experimental data with theoretical results is of particular interest for proving a better
 understanding of these materials.  
 
 Several theoretical studies have been performed previously on these compounds, most of them based on pair-wise interatomic potentials \cite{Shchur1999,Shchur2006}. 
 From a Density Functional Theory point of view, the studies are quite recent \cite{Shchur2016,Lasave2016}. Some\cite{Shchur2016} pointed out that the Local-Density Approximation (LDA)\cite{Martin2004} is
 not able to properly describe CDP, predicting the $P2_1/m$ phase to be more stable than the $P2_1$ one at 0K, 
 in contradiction with experiments. Similar problems are encountered when the exchange-correlation is approximated by the Perdew-Burke-Ernzerhof (PBE) formulation \cite{Perdew1996}
 of the Generalized-Gradient Approximation (GGA) \cite{Martin2004}. GGA-PBE, indeed, correctly predicts the ferroelectric phase to be energetically favorable than the paraelectric one,
but it largely overestimates the lattice parameters (about 8\%, as shown later).
 
 Interestingly, the breakdown of these commonly-used (semi-)local DFT exchange-correlation functionals has not been encountered for other hydrogen-based 
 ferroelectrics like KH$_2$PO$_4$ \cite{Zhang2001} or NH$_4$H$_2$PO$_4$ \cite{Lasave2007}, for which experimental lattice parameters are reproduced within a few percent using PBE.
 As discussed in the body of this paper, we suspect that, while the LDA failure can be attributed mainly to the incorrect description of hydrogen bonds, the one of PBE derives
 from the incorrect description of the long-range e$^-$-e$^-$ correlation, which plays an important role in Cs-based compounds due to the large polarizability of the Cs ion. 
 This assumption is supported by the inability of PBE to predict the correct stable phase of CsCl, CsBr and CsI as
 shown by Zhang \textit{et al.} \cite{Zhang2013}.  
The importance of the dispersion corrections was also highlighted in the case of CsH$_2$SO$_4$ \cite{Krzystyniak2015} and
 for Cs - Cs and Na - Cs alkali-metal dimers by Ferri \textit{et al.} \cite{Ferri2015}.

 In consequence, special care needs to be taken when choosing the exchange-correlation and the role of the dispersion corrections has to be examined carefully. Several techniques have been proposed in the recent years for the
 description of long-range correlation, including vdW-DF \cite{Dion2004, Lee2010}, TS-vdW \cite{Tkatchenko2009}, MBD \cite{Reilly2013} or Grimme's DFT-D methods \cite{Grimme2006,Grimme2010,Grimme2010b}. The performances of
 some of these techniques have already been investigated in the case of hydrogen bonds \cite{Maron2011} revealing that TS-vdW performs usually better in molecules (test set S22), while vdW-DF2 and DFT-D3 give similar precision for this test set. 
 On contrary, their performances for hydrogen-based crystals have not been yet systematically investigated to the best of our knowledge.  
 
 In this work, we chose to compare the performance of DFT-D3 \cite{Grimme2010} and DFT-D3(BJ) \cite{Grimme2010b} methods, motivated in part
  by the implementation within DFPT of the DFT-D contributions to the vibrational properties \cite{Gonze1997,Gonze1997b,Hamann2005,Hamann2005b,Vantroeye2016}. 
 This formalism allows for the efficient computation of response function properties including 
 dielectric constants, Born effective charges and phonon band structures, that would otherwise require the use of supercells for describing the
 collective atomic motion characterized by wavevectors in the whole Brillouin zone, or an homogeneous electric field not compatible with periodic boundary conditions.
 We also extend this formalism to include the DFT-D contributions to strain-related properties e.g. elastic constants, which were unavailable to our knowledge in DFPT up to now.
 
 Thanks to these DFT-D corrections, we are able to properly reproduce both the correct ground-state of CDP and DCDP together with their vibrational properties. We also predict
 the elastic and piezoelectric constants of this material, as well as its spontaneous polarization using the Berry phase technique.
 
  This manuscript is organized as follows: in Sec. \ref{Crystal}, the CDP crystal structure and the computational methods are detailed.
  In Sec. \ref{Results}, the structural, dielectric, vibrational and elastic properties of both CDP and DCDP are presented as well as their phase stability.
  We detail the theoretical derivations of the dispersion contributions to the elastic constants in Appendix \ref{Impl} and
  validate the correctness of the implementation with respect to finite differences.

\section{Crystal structure and computational method} \label{Crystal}
All computations are performed with the \textsc{Abinit} software \cite{Abinit2005,Abinit2009,Gonze2016}. The exchange-correlation energy is approximated using the GGA-PBE functional.
Dispersion corrections obtained
by DFT-D3 or DFT-D3(BJ) are also considered. For the sake of brevity, uncorrected GGA-PBE are denoted as PBE and the results with dispersion corrections by DFT-D3 and DFT-D3(BJ) as PBE-D3 and PBE-D3(BJ), respectively.
The cut-off radius for the coordination number, required for the vdW corrections, is set 
to 105~\AA $\,$  and only pairs contributing for more than $10^{-12}$ Ha are taken into account. 
We use ONCVPSP norm-conserving pseudopotentials \cite{Hamann2013} which include
multiple projectors to treat semi-core and unbound states. The corresponding average $\Delta$-gauge \cite{Lejaeghere2016} is about 0.3 meV.
These pseudopotentials are presented in more details in the Supplemental Material \footnote{See Supplemental Material at ... for additional information about the pseudopotentials, lattice parameters
of the studied hydrogen-based ferroelectrics, reduced coordinates of the P2$_1$ (Z=2) and P2$_1$ (Z=4) phases, as well as the phonon frequencies of the P2$_1$ (Z=4) phase.}.

 
The geometries of the CDP phases are relaxed until the forces are all smaller than 10$^{-8}$ Ha/Bohr.
All other properties (i.e. interatomic force constants, elastic tensor, etc.) are computed at the relaxed lattice parameters.
A plane-wave cut-off energy of 60 Ha and a $4\times4\times4$ Monkhorst-Pack grid of k-points \cite{Monkhorst1976} are found sufficient to obtain a precision better than
0.5 mHa/atom on the total energy, 0.2\% on the lattice parameters, and $1$ cm$^{-1}$ on the phonon frequencies at the Brillouin-zone center. 
In all the computations, we use an energy cut-off smearing of 0.5 Ha \cite{Bernasconi1995,Janssen2016}. 

First-order response functions are computed within DFPT, including the contributions of DFT-D methods as discussed in Ref. \onlinecite{Vantroeye2016} and in the appendix of the present paper. 
For all the previously-mentioned properties, we neglect the effect of zero-point motion.

\begin{figure}[h]
\hspace{-0.2cm}\includegraphics[width=0.25\textwidth]{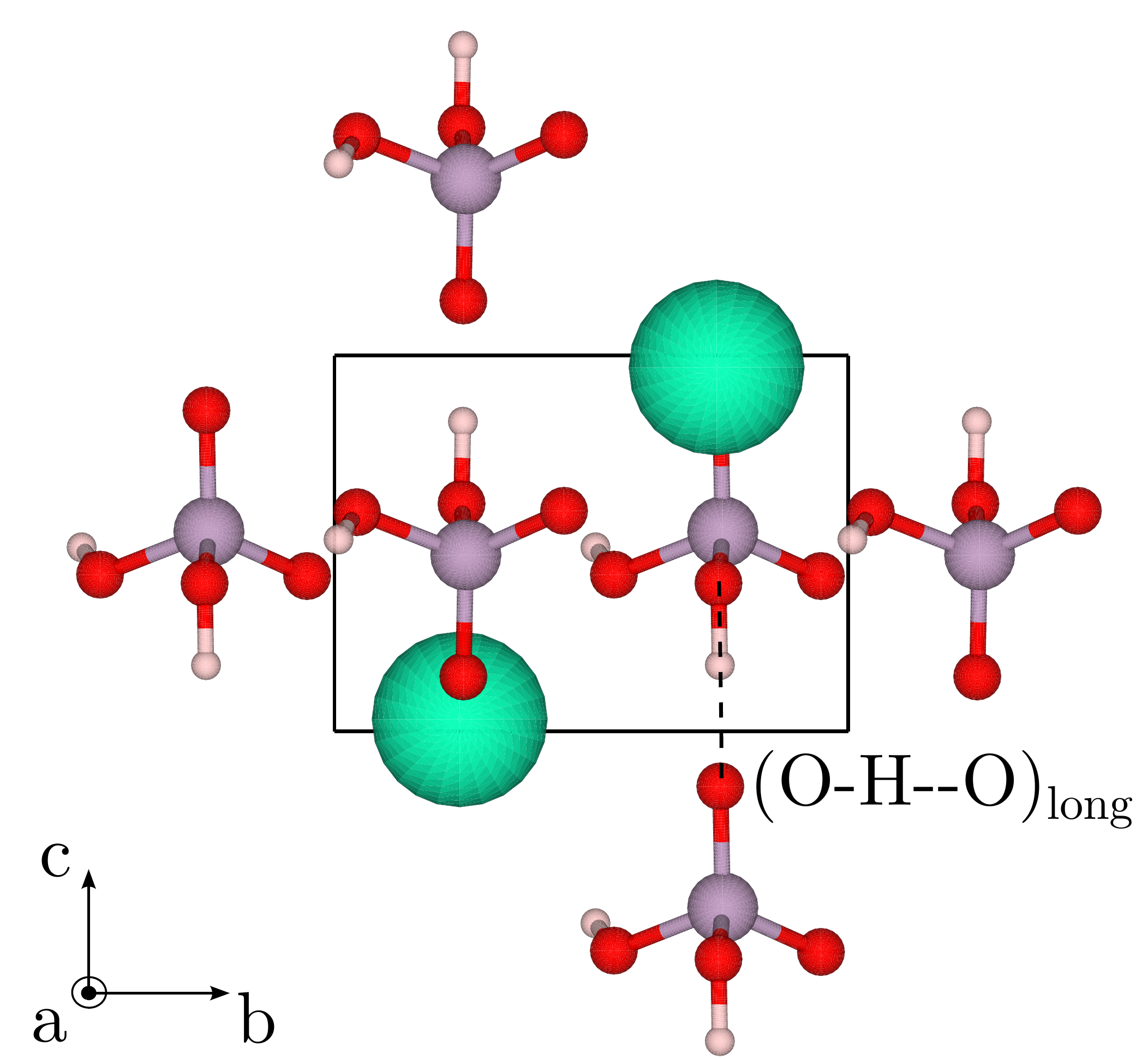}
\hspace{-0.3cm}\includegraphics[width=0.25\textwidth]{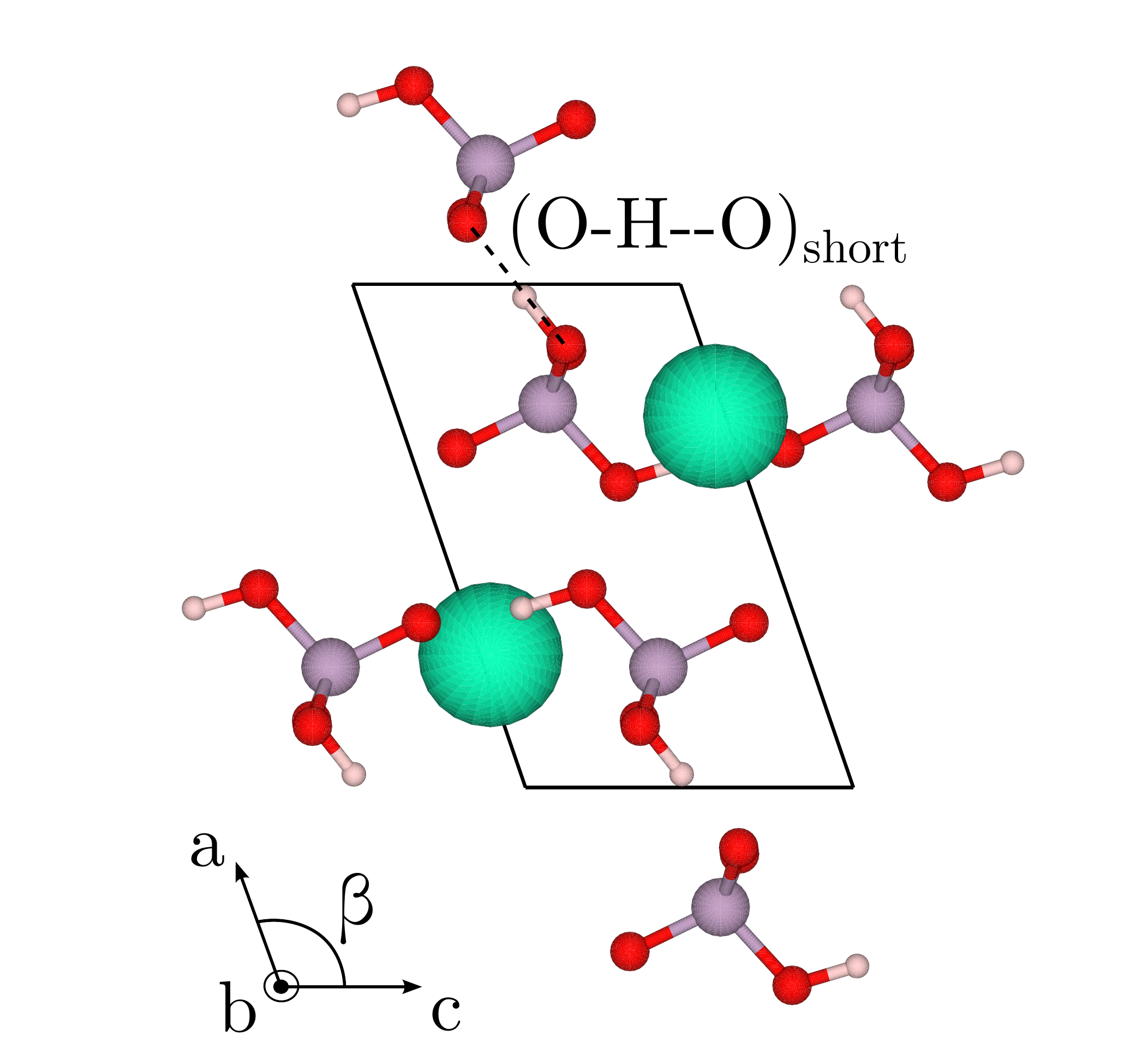}
\begin{center}
 (a) $P2_1$ ferroelectric phase
\end{center}
\hspace{-0.2cm}\includegraphics[width=0.25\textwidth]{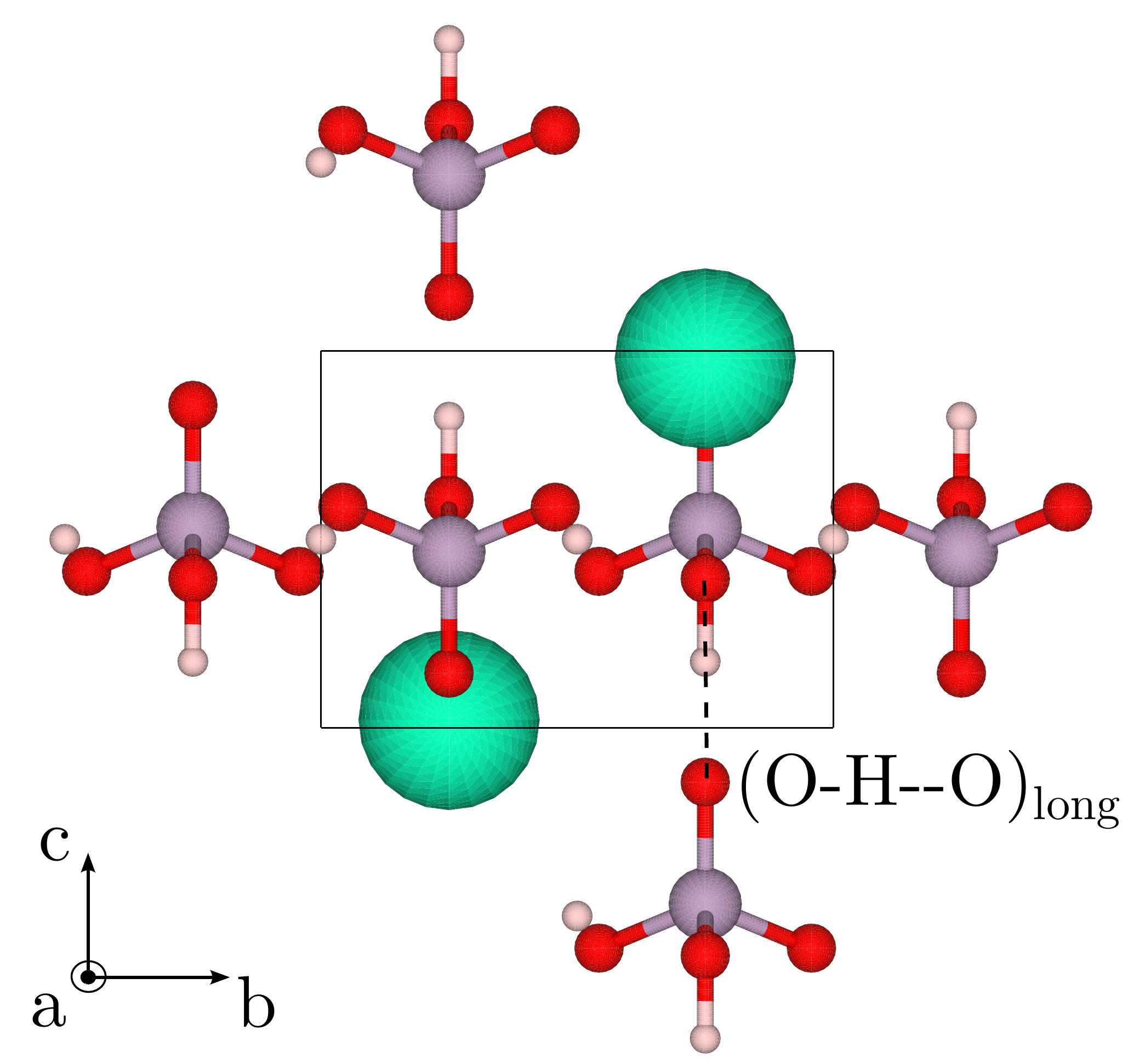}
\hspace{-0.3cm}\includegraphics[width=0.25\textwidth]{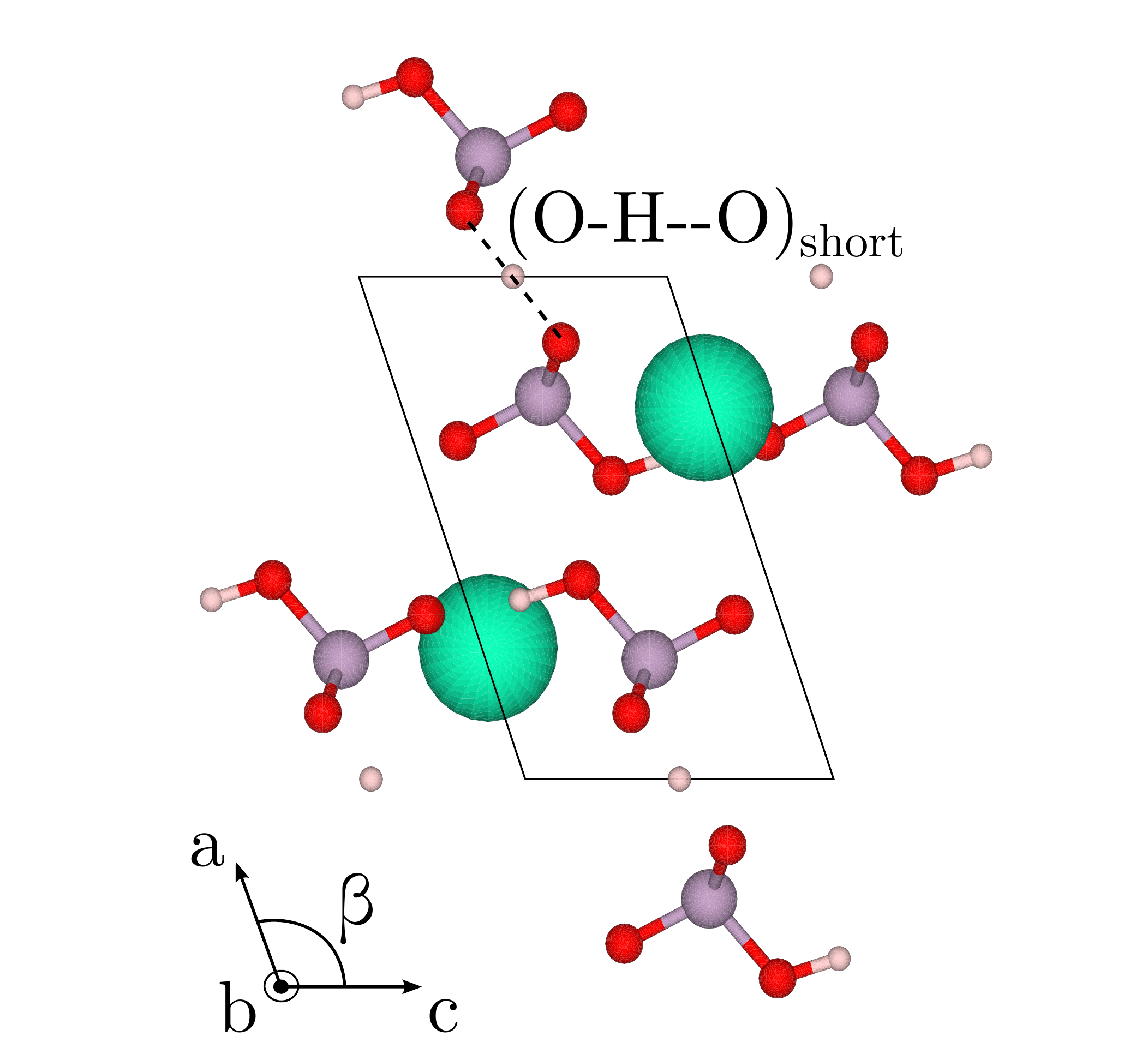}
\begin{center}
 (b) $P2_1/m$ paraelectric phase
\end{center}
\caption{\label{Graphics1} \textsc{Color online.} Primitive cell of CDP (DCDP) crystal in (a) the ferroelectric phase (b) the paraelectric phase. The Cs, P, O and H(D)
atoms are represented in green, purple, red and white, respectively. The two inequivalent hydrogen bonds (O-H---O)$_{\text{long}}$ and (O-H---O)$_{\text{short}}$ are represented with dashed line. 
These figures were generated with the \textsc{Vesta} software \cite{Momma2011}.}
\end{figure}

CDP and DCDP crystallize at low temperature into a monoclinic structure ($P2_1$ symmetry, $Z=2$). The corresponding primitive cell of CDP is depicted in~Fig. \ref{Graphics1}. 
The $b$ axis is chosen as the unique axis. The hydrogenated phosphate groups of a given CDP unit cell form hydrogen bonds with
those situated in the neighboring cells as shown in Fig. \ref{Graphics1}. However, these hydrogen bonds are inequivalent: the one made along the $c$ axis, denoted (O-H---O)$_{\text{long}}$, is
shorter than the one along $b$ axis, denoted (O-H---O)$_{\text{short}}$. 
At higher temperature, this ferroelectric crystal undergoes a phase transition to a $P2_1/m$ paraelectric phase ($Z=2$), involving the motion of the hydrogen atom involved in (O-H---O)$_{\text{short}}$
 into a median position between the O atoms. Apart from this displacement, the two phases are otherwise similar, as illustrated in Fig.~\ref{Graphics1}.

\section{Ground-state} \label{Results}

In this section, we report the computed geometry for both the ferroelectric and the paraelectric phases of CDP. The computed relaxed lattice parameters ($a$,$b$,$c$) for both phases are summarized in Table \ref{abcFerro},
alongside with the corresponding available experimental data \cite{Frazer1979}. As we neglect the zero-point motion, the structure of DCDP is equivalent to the one of CDP. As the experimentally-reported
thermal expansion of CDP is extremely small~\cite{Nakamura1984}, leading to changes of approximatively $ -4$~m\AA $\,$ when the (experimental) lattice parameter $b$ was extrapolated to 0K for the ferroelectric phase, we will neglect
thermal expansion in the following of the paper.

\begin{table}[h]
\begin{center}
 \begin{tabular}{l c c c c}
 \hline \hline
 & $a$ [\AA] & $b$ [\AA] & $c$ [\AA] & $\beta$ [$^\circ$] \\
 \hline
 \multicolumn{4}{l}{Ferroelectric phase} \\
 \hline
  PBE & 8.50 & 6.48 & 4.97 & 112.0 \\
  PBE-D3 & 7.92 & 6.16 &4.91  &109.8\\
  PBE-D3(BJ) & 7.98 & 6.36 & 4.91& 109.0\\
  Exp.$^a$ & 7.899 & 6.294 & 4.898 & 108.50\\
  \hline
  \multicolumn{4}{l}{Paraelectric phase} \\
 \hline
  PBE & 8.18 & 6.61 & 4.97 & 110.0 \\
  PBE-D3 & 7.84 & 6.15 &4.90  &109.3 \\
  PBE-D3(BJ) & 7.92 & 6.34 & 4.92& 108.7\\ 
  Exp.$^b$ & 7.899 & 6.325 & 4.890 & 108.29\\
  \hline \hline
  $^a$ 100K, Ref. \onlinecite{Frazer1979} \\
  $^b$ 200K, Ref. \onlinecite{Frazer1979}
 \end{tabular}
 \caption{Lattice parameters of ferroelectric and paraelectric CDP computed with the different functionals, as well as the experimental results from Frazer \textit{et al.}\cite{Frazer1979}. \label{abcFerro}}
\end{center}
\end{table}

Previous ab-initio calculations using LDA \cite{Shchur2016} did not find the ferroelectric phase to be the most stable at low temperature in disagreement with experiments. This failure is not surprising, as on average LDA 
underperforms in the description of hydrogen bonds\cite{Maron2011}. In contrast, PBE correctly predicts the ferroelectric phase to be more stable than the paraelectric one. However, the theoretical value of 8.50~\AA \
for the lattice parameter $a$ overestimates the experimental value by more than 8\%. This is quite unexpected, given that
no particular problems were reported for other hydrogen-based ferroelectrics \cite{Zhang2001,Lasave2007}. Interestingly, in the case of CsHSO$_4$ \cite{Krzystyniak2015}, which crystallizes at low temperature
into a $P21/c$ structure (Z=4), PBE also shows some failures. On the one hand, it strongly overestimates the experimental unit cell volume; and, on the other hand, it leads to lattice instabilities \footnote{See supplementary information of Ref.\onlinecite{Krzystyniak2015}}.
This problem was overcome when vdW corrections were added to PBE or rPBE. Similarly, DFT-D2 vdW corrections were required to properly describe the ground-state of Cs-halogens starting from PBE \cite{Zhang2013}.

When vdW corrections are included, we are able to reach a good agreement with the experiments in the case of CDP 
[3\% maximum relative error for PBE-D3, even better with PBE-D3(BJ)]. 
In order to investigate further the correctness of these DFT-D corrections in the case of hydrogen-based ferroelectrics, we compute the relaxed structures of KH$_2$PO$_4$ and RbH$_2$PO$_4$ 
for both the ferroelectric and the paraelectric phases and compare the predictions with experiments.
We find that the average relative errors of PBE, PBE-D3 and PBE-D3(BJ) are 2.8\%, 1.4\% and 0.5\%, respectively. This indicates that PBE-D3 and PBE-D3(BJ) are able to describe these materials at least in terms of structures. 
One should mention though, that the most marked effects were observed for CDP, which contains highly polarizable Cs-atoms. More details can be found in the Supplemental Material.
Since the PBE functional fails to predict the geometry of these hydrogen-based ferroelectric materials, it was discarded for the subsequent computations.

Concerning the (O-H--O) bonds, PBE-D3 and PBE-D3(BJ) give similar values for both the short and the long bond lengths (see Fig. \ref{Graphics1}), i.e. 2.49~\AA \, and 2.52~\AA \, for PBE-D3 and 2.49~\AA \, and 2.535 \AA \, for PBE-D3(BJ). These values should
be compared to the experimental bonds i.e. 2.48~\AA \, and 2.54~\AA\cite{Frazer1979}. In view of these results, PBE-D3(BJ) performs slightly better for these bond lengths.

\subsection{Stability}

In order to investigate the stability of the ferroelectric phase of CDP (DCDP) compared to the paraelectric one, we have first computed the energy difference per atom between these two phases $\Delta E = E_{P2_1}-E_{P2_1/m}$.
For PBE-D3 and PBE-D3(BJ), the results are  -3.1 and -2.9~meV/atom, respectively. In agreement with the experiments, the ferroelectric
phase is found to be more stable than the paraelectric one. 

\begin{figure}[ht]
\begin{center}
    \begingroup%
  \makeatletter%
  \providecommand\color[2][]{%
    \errmessage{(Inkscape) Color is used for the text in Inkscape, but the package 'color.sty' is not loaded}%
    \renewcommand\color[2][]{}%
  }%
  \providecommand\transparent[1]{%
    \errmessage{(Inkscape) Transparency is used (non-zero) for the text in Inkscape, but the package 'transparent.sty' is not loaded}%
    \renewcommand\transparent[1]{}%
  }%
  \providecommand\rotatebox[2]{#2}%
  \ifx\svgwidth\undefined%
    \setlength{\unitlength}{285bp}%
    \ifx\svgscale\undefined%
      \relax%
    \else%
      \setlength{\unitlength}{\unitlength * \real{\svgscale}}%
    \fi%
  \else%
    \setlength{\unitlength}{\svgwidth}%
  \fi%
  \global\let\svgwidth\undefined%
  \global\let\svgscale\undefined%
  \makeatother%
  
  \begin{picture}(1,0.64384615)%
    \put(-0.07,0){\includegraphics[width=\unitlength]{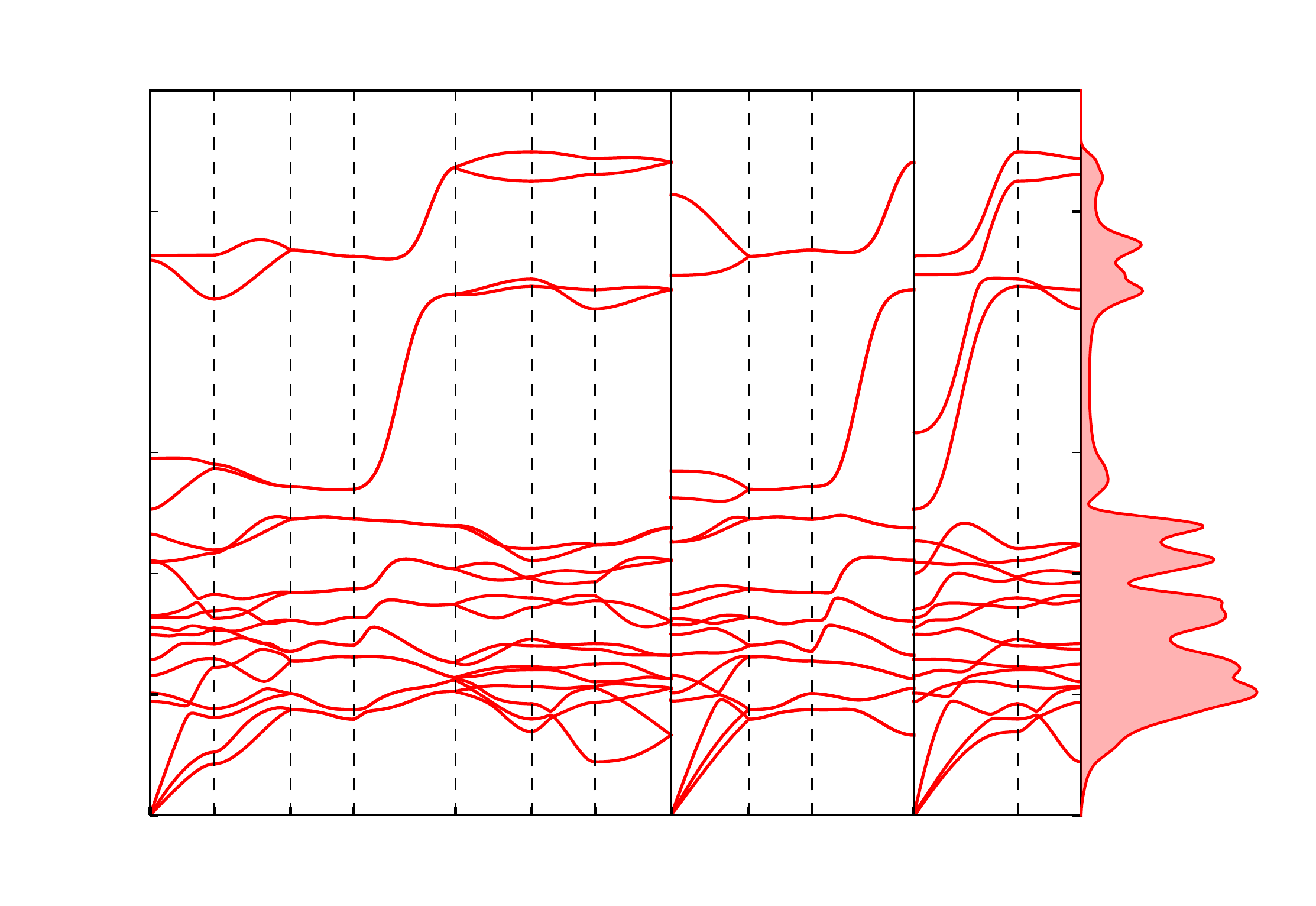}}%
    \put(0.034816239,0.03752457){\makebox(0,0)[lb]{\smash{$\Gamma$}}}%
    \put(0.0803227067,0.03752457){\makebox(0,0)[lb]{\smash{Y}}}%
    \put(0.14205373,0.03752457){\makebox(0,0)[lb]{\smash{C}}}%
    \put(0.189450263,0.03752457){\makebox(0,0)[lb]{\smash{Z}}}%
    \put(0.26730868,0.03752457){\makebox(0,0)[lb]{\smash{D}}}%
    \put(0.32156431,0.03752457){\makebox(0,0)[lb]{\smash{B}}}%
    \put(0.362257465,0.03752457){\makebox(0,0)[lb]{\smash{A$_0$}}}%
    \put(0.40667556,0.03752457){\makebox(0,0)[lb]{\smash{E$_0|\Gamma$}}}%
    \put(0.49159325,0.03752457){\makebox(0,0)[lb]{\smash{Z}}}%
    \put(0.5350717571,0.03752457){\makebox(0,0)[lb]{\smash{C}}}%
    \put(0.595388915,0.03752457){\makebox(0,0)[lb]{\smash{E$|\Gamma$}}}%
    \put(0.69216915,0.03752457){\makebox(0,0)[lb]{\smash{B}}}%
    \put(0.737314904,0.03752457){\makebox(0,0)[lb]{\smash{A}}}%
    \put(0.020782585,0.06791934){\makebox(0,0)[lb]{\smash{0}}}%
    \put(0.003191774,0.14843216){\makebox(0,0)[lb]{\smash{50}}}%
    \put(-0.01191774,0.24394497){\makebox(0,0)[lb]{\smash{100}}}%
    \put(-0.01191774,0.33445779){\makebox(0,0)[lb]{\smash{150}}}%
    \put(-0.01191774,0.42797062){\makebox(0,0)[lb]{\smash{200}}}%
    \put(-0.01191774,0.51748344){\makebox(0,0)[lb]{\smash{250}}}%
    \put(-0.01191774,0.61099626){\makebox(0,0)[lb]{\smash{300}}}%
    \put(0.06191774,0.6291099626){\makebox(0,0)[lb]{\smash{Frequency [cm$^{-1}$]}}}%
\put(0.403191774,-0.01099626){\makebox(0,0)[lb]{\smash{(a)}}}%
  \end{picture}%
\endgroup%
    
    \vspace{0.7cm}
   \begingroup%
  \makeatletter%
  \providecommand\color[2][]{%
    \errmessage{(Inkscape) Color is used for the text in Inkscape, but the package 'color.sty' is not loaded}%
    \renewcommand\color[2][]{}%
  }%
  \providecommand\transparent[1]{%
    \errmessage{(Inkscape) Transparency is used (non-zero) for the text in Inkscape, but the package 'transparent.sty' is not loaded}%
    \renewcommand\transparent[1]{}%
  }%
  \providecommand\rotatebox[2]{#2}%
  \ifx\svgwidth\undefined%
    \setlength{\unitlength}{285bp}%
    \ifx\svgscale\undefined%
      \relax%
    \else%
      \setlength{\unitlength}{\unitlength * \real{\svgscale}}%
    \fi%
  \else%
    \setlength{\unitlength}{\svgwidth}%
  \fi%
  \global\let\svgwidth\undefined%
  \global\let\svgscale\undefined%
  \makeatother%
  
  \hspace*{0.05cm} \begin{picture}(1,0.64384615)%
    \put(-0.07,0){\includegraphics[width=\unitlength]{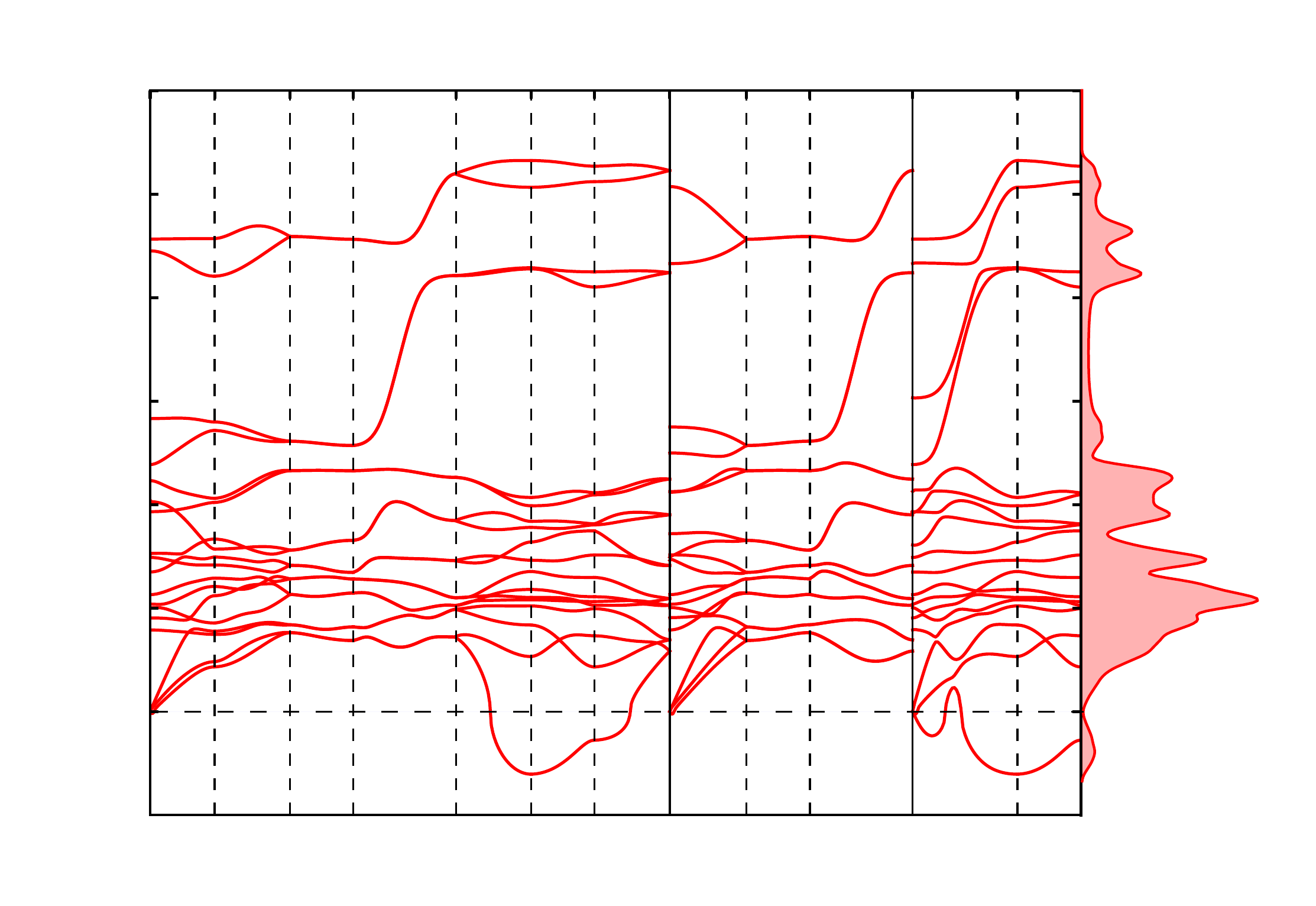}}%
    \put(0.044816239,0.03752457){\makebox(0,0)[lb]{\smash{$\Gamma$}}}%
    \put(0.0803227067,0.03752457){\makebox(0,0)[lb]{\smash{Y}}}%
    \put(0.14205373,0.03752457){\makebox(0,0)[lb]{\smash{C}}}%
    \put(0.189450263,0.03752457){\makebox(0,0)[lb]{\smash{Z}}}%
    \put(0.26730868,0.03752457){\makebox(0,0)[lb]{\smash{D}}}%
    \put(0.32156431,0.03752457){\makebox(0,0)[lb]{\smash{B}}}%
    \put(0.362257465,0.03752457){\makebox(0,0)[lb]{\smash{A$_0$}}}%
    \put(0.40667556,0.03752457){\makebox(0,0)[lb]{\smash{E$_0|\Gamma$}}}%
    \put(0.49159325,0.03752457){\makebox(0,0)[lb]{\smash{Z}}}%
    \put(0.5350717571,0.03752457){\makebox(0,0)[lb]{\smash{C}}}%
    \put(0.595388915,0.03752457){\makebox(0,0)[lb]{\smash{E$|\Gamma$}}}%
    \put(0.69216915,0.03752457){\makebox(0,0)[lb]{\smash{B}}}%
    \put(0.737314904,0.03752457){\makebox(0,0)[lb]{\smash{A}}}%
    \put(-0.007191774,0.06791934){\makebox(0,0)[lb]{\smash{-50}}}%
    \put(0.020782585,0.137243216){\makebox(0,0)[lb]{\smash{0}}}%
    \put(0.003191774,0.217394497){\makebox(0,0)[lb]{\smash{50}}}%
    \put(-0.01191774,0.29445779){\makebox(0,0)[lb]{\smash{100}}}%
    \put(-0.01191774,0.37297062){\makebox(0,0)[lb]{\smash{150}}}%
    \put(-0.01191774,0.44748344){\makebox(0,0)[lb]{\smash{200}}}%
    \put(-0.01191774,0.53099626){\makebox(0,0)[lb]{\smash{250}}}%
    \put(-0.01191774,0.61099626){\makebox(0,0)[lb]{\smash{300}}}%
    \put(0.06191774,0.6291099626){\makebox(0,0)[lb]{\smash{Frequency [cm$^{-1}$]}}}%
\put(0.403191774,-0.01099626){\makebox(0,0)[lb]{\smash{(b)}}}%
  \end{picture}%
\endgroup%

  \caption{Low-frequency phonon band structure of $P2_1$ CDP computed with (a) PBE-D3 and (b) PBE-D3(BJ) as well as the corresponding phonon density of states.
    \label{PhononStruc-D3}}

\end{center}
\end{figure}
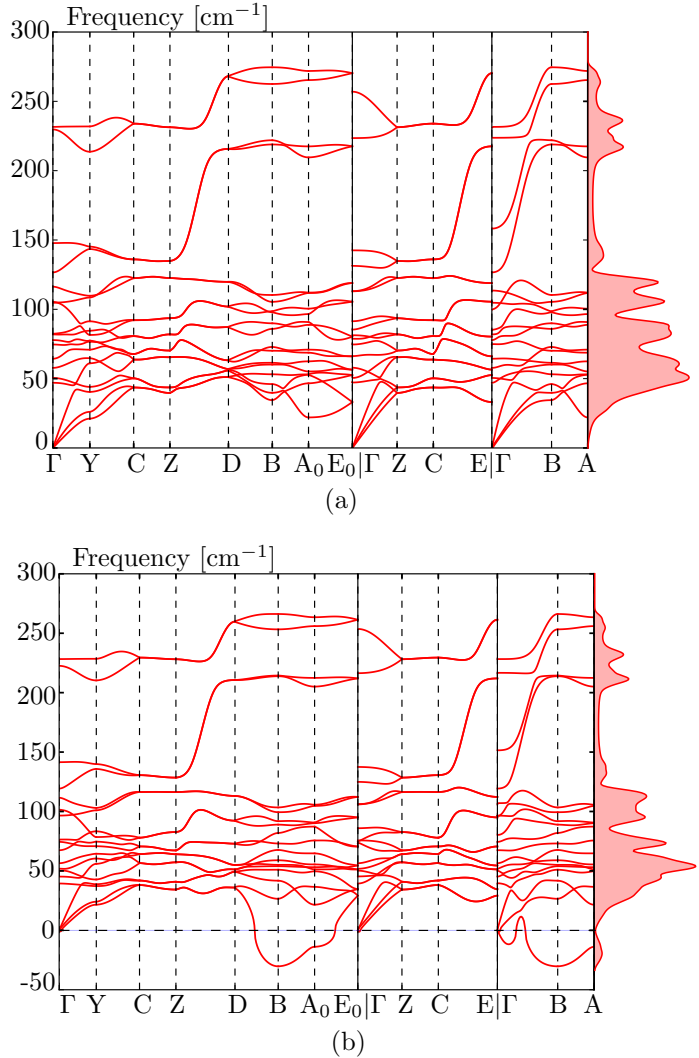

However, this does not guarantee that the ferroelectric phase is the global energy minimum.
In order to consider also the dynamical stability, the phonon band structures have been computed for the ferroelectric phase of CDP with PBE-D3 and
PBE-D3(BJ). In these calculations, we have used a $4\times4\times4$ and $2\times2\times2$ phonon wavector supporting mesh for the interpolation technique described in Ref. \onlinecite{Gonze1997}.
The results are presented in Fig. \ref{PhononStruc-D3}. The trends are similar for DCDP (not presented).

The PBE-D3 phonon band structure shows only positive frequencies. Hence, the $P2_1$ structure is stable at 0K. In contrast, the PBE-D3(BJ) phonon band structure presents instabilities around the $B$ and $A_{0}$ high
symmetry points indicating the existence of a lower-energy phase.

\begin{figure}[ht]
\hspace{-0.2cm}\includegraphics[width=0.25\textwidth]{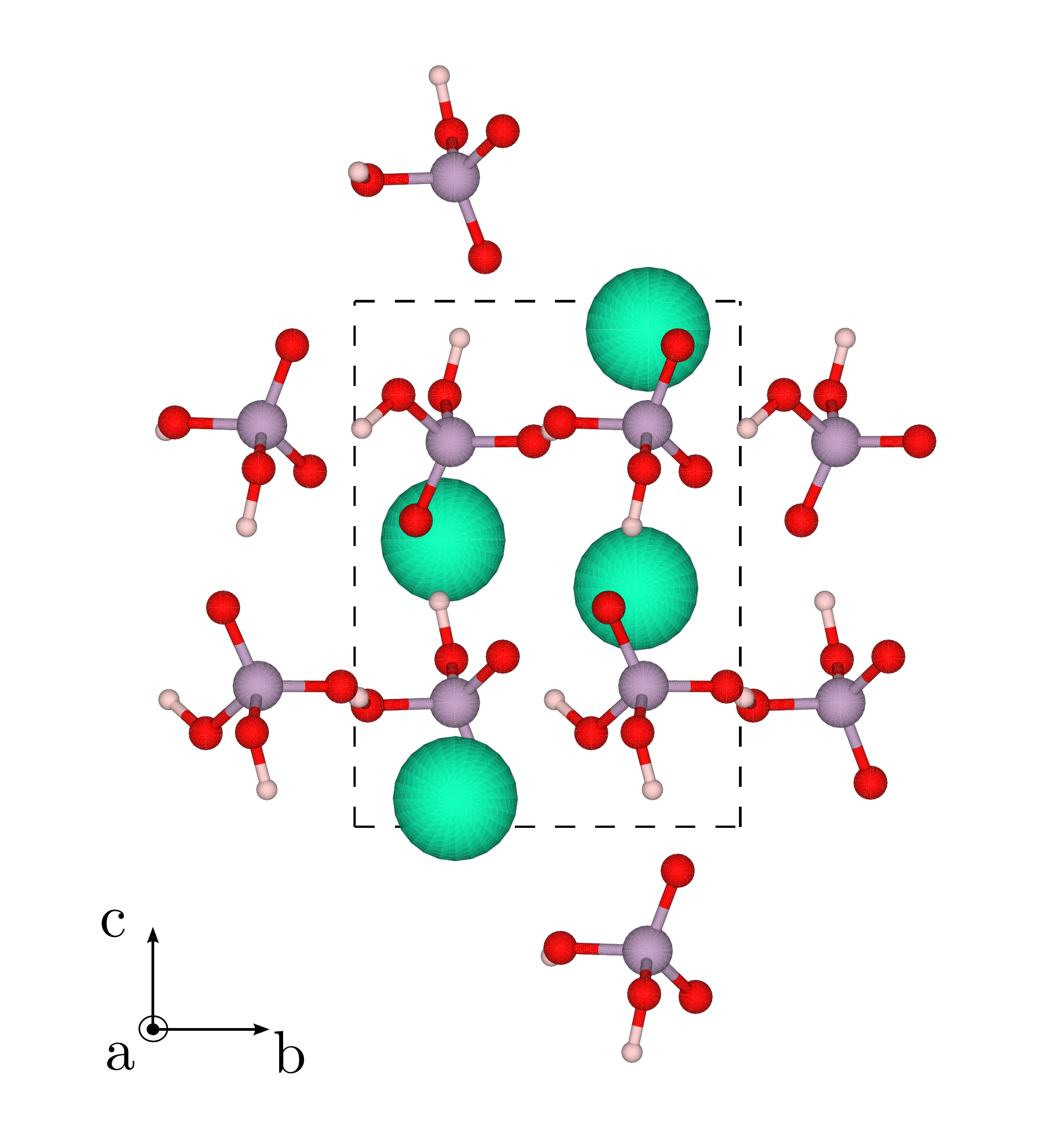}
\hspace{-0.3cm}\includegraphics[width=0.25\textwidth]{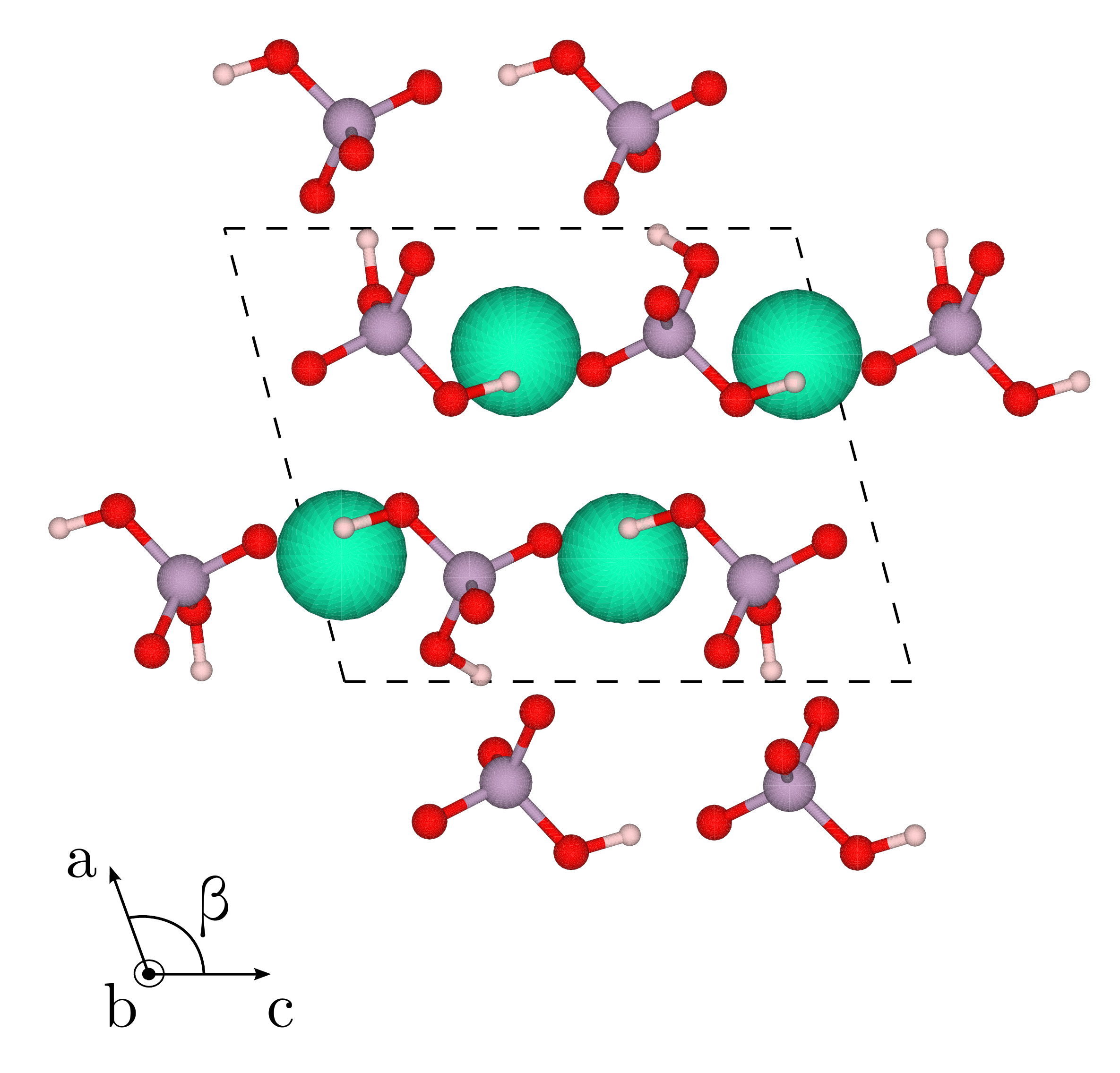}
\caption{\label{Graphics} \textsc{Color online.} Primitive cell of the 0K stable structure predicted by PBE-D3(BJ) and the corresponding phonon density of states. The cyan atoms correspond to Cs, the purple ones to P, the red ones to O and the white ones to H
(D in case of deuterium). \label{SuperCrystal}}
\end{figure}

In order to investigate further this instability, we have considered a $P2_1$ 1$\times$1$\times$2 supercell. We have moved the atoms according to the previously-found
unstable phonon mode at the B high-symmetry point (where the strongest instability was found) and then fully relaxed the atomic positions and lattice parameters.
The corresponding relaxed structure -that preserves the
$P2_1$ symmetry- is shown in Fig. \ref{SuperCrystal}. Its predicted lattice parameters are $a$=7.79~\AA,  $b$=6.71~\AA , $c$=9.47~\AA $\,$ and $\beta$=104.9$^\circ$. 
The (O-H--O) bonds are different from the ones in the $P2_1$ (Z=2) phase, for which the short (O-H--O) bonds are 2.49~\AA \, long and the long (O-H--O) bonds are 2.535~\AA \, long.  
In the new structure, these values become two different values both for the former (2.50~\AA \, and 2.51~\AA) and for
the latter (2.49~\AA \, and 2.52~\AA). Consequently, the distinction between long (O-H--O) bond along the $a$ axis and short (O-H--O) bond along the $c$ axis becomes less appropriate for this new phase compared to the $P2_1$ (Z=2) 
phase. The reduced coordinates are reported in the Supplemental Material. It lies
-3.3~meV/atom below the experimentally-observed ferroelectric structure in PBE-D3(BJ), while it is simply unstable in PBE-D3. 

\begin{figure}[ht]
\begin{center}
    \begingroup%
  \makeatletter%
  \providecommand\color[2][]{%
    \errmessage{(Inkscape) Color is used for the text in Inkscape, but the package 'color.sty' is not loaded}%
    \renewcommand\color[2][]{}%
  }%
  \providecommand\transparent[1]{%
    \errmessage{(Inkscape) Transparency is used (non-zero) for the text in Inkscape, but the package 'transparent.sty' is not loaded}%
    \renewcommand\transparent[1]{}%
  }%
  \providecommand\rotatebox[2]{#2}%
  \ifx\svgwidth\undefined%
    \setlength{\unitlength}{285bp}%
    \ifx\svgscale\undefined%
      \relax%
    \else%
      \setlength{\unitlength}{\unitlength * \real{\svgscale}}%
    \fi%
  \else%
    \setlength{\unitlength}{\svgwidth}%
  \fi%
  \global\let\svgwidth\undefined%
  \global\let\svgscale\undefined%
  \makeatother%
  
    \begin{picture}(1,0.64384615)%
    \put(-0.07,0){\includegraphics[width=\unitlength]{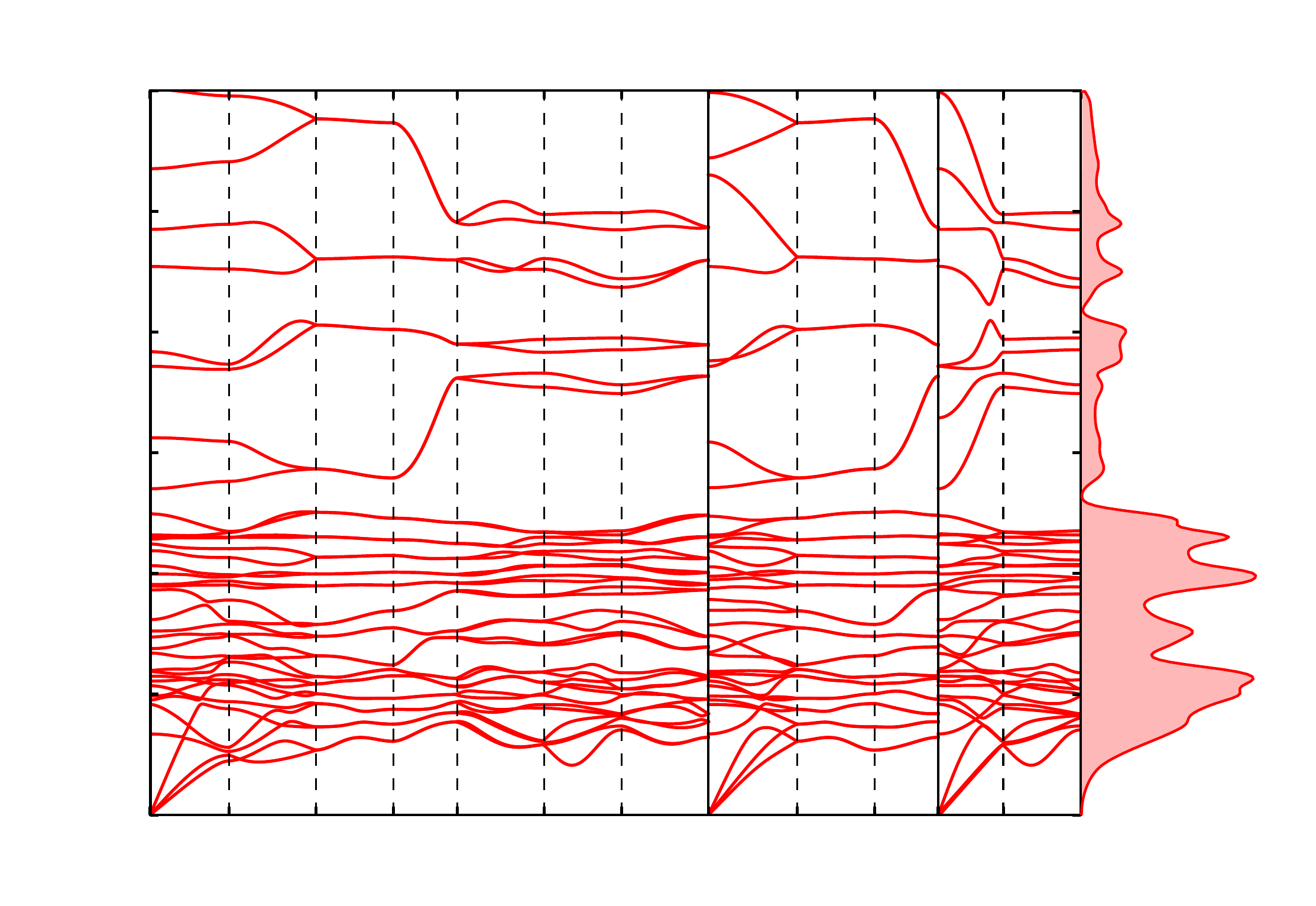}}%
    \put(0.034816239,0.03752457){\makebox(0,0)[lb]{\smash{$\Gamma$}}}%
    \put(0.0923227067,0.03752457){\makebox(0,0)[lb]{\smash{Y}}}%
    \put(0.157505373,0.03752457){\makebox(0,0)[lb]{\smash{C}}}%
    \put(0.2161450263,0.03752457){\makebox(0,0)[lb]{\smash{Z}}}%
    \put(0.27430868,0.03752457){\makebox(0,0)[lb]{\smash{D}}}%
    \put(0.3367156431,0.03752457){\makebox(0,0)[lb]{\smash{B}}}%
    \put(0.3852257465,0.03752457){\makebox(0,0)[lb]{\smash{A$_0$}}}%
    \put(0.431667556,0.03752457){\makebox(0,0)[lb]{\smash{E$_0|\Gamma$}}}%
    \put(0.525159325,0.03752457){\makebox(0,0)[lb]{\smash{Z}}}%
    \put(0.5800717571,0.03752457){\makebox(0,0)[lb]{\smash{C}}}%
    \put(0.615388915,0.03752457){\makebox(0,0)[lb]{\smash{E$|\Gamma$}}}%
    \put(0.68216915,0.03752457){\makebox(0,0)[lb]{\smash{B}}}%
    \put(0.737314904,0.03752457){\makebox(0,0)[lb]{\smash{A}}}%
    \put(0.020782585,0.06791934){\makebox(0,0)[lb]{\smash{0}}}%
    \put(0.003191774,0.14843216){\makebox(0,0)[lb]{\smash{50}}}%
    \put(-0.01191774,0.24394497){\makebox(0,0)[lb]{\smash{100}}}%
    \put(-0.01191774,0.33445779){\makebox(0,0)[lb]{\smash{150}}}%
    \put(-0.01191774,0.42797062){\makebox(0,0)[lb]{\smash{200}}}%
    \put(-0.01191774,0.51748344){\makebox(0,0)[lb]{\smash{250}}}%
    \put(-0.01191774,0.61099626){\makebox(0,0)[lb]{\smash{300}}}%
    \put(0.06191774,0.6291099626){\makebox(0,0)[lb]{\smash{Frequency [cm$^{-1}$]}}}%
  \end{picture}%
\endgroup%
    \vspace{-0.8cm} \caption{Low-frequency phonon band structure of the predicted stable phase of PBE-D3(BJ) and the corresponding phonon density of states. These
    results are obtained by interpolating the ab-initio dynamical matrices computed on a $4\times4\times2$ phonon wavector mesh following Ref. \onlinecite{Gonze1997}.
  \label{PhononStruc-SC}}
\end{center}
\end{figure}
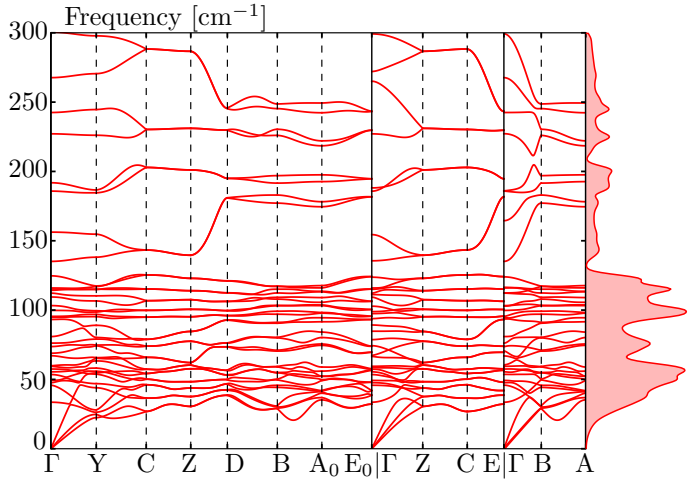

We have also computed the phonon band structure of this new phase, which does not show any structural instability (see Fig. \ref{PhononStruc-SC}). 
Interestingly, the corresponding lattice phonon density of states shows a large amount of similarities with respect to the one obtained with PBE-D3 for the ferroelectric phase. The phonon analysis of
this phase is shown in the Supplemental Material.

To the best of our knowledge, temperature measurements have not been performed under 80K for this material. Consequently the existence of a low temperature phase can not be discarded. However,
one should be cautious with respect to the PBE-D3(BJ) results. On the one hand, we are neglecting the quantum nature of the nuclei, which is crucial in the case of hydrogen-based
compounds. On the other hand,
the existence of this phase could be a spurious effect of the approximate long-range e$^-$-e$^-$ correlation in use. For these reasons, we have decided not to consider PBE-D3(BJ) for the remaining studies of the response-function properties.

\subsection{Vibrational analysis}

The phonon modes of the $P2_1$ (Z=2) ferroelectric phase of CDP (DCDP) consists of 3 acoustic modes, 15 lattice modes (8A$\oplus$7B) i.e. intermolecular vibrations and 30 internal modes (15A$\oplus$15B).
The isotopic substitution (hydrogen or deuterium) will mostly affect the eigenfrequencies of the modes that are dominated by hydrogen (deuterium) motion. 
If there were modes with only hydrogen (deuterium) motion, the ratio of frequencies would be the square root of their mass ratio, i.e. $\sqrt{2} \approx$1.414.

The phonon frequencies of ferroelectric CDP and DCDP at $\Gamma$ 
for the A modes and B modes can be found in Tables \ref{PhononCDP} and \ref{PhononCDP2}, respectively. The experimental results from Marchon and Novak \cite{Marchon1983} (Raman and infrared reflectivity) 
and the corresponding mode assignments based on our analysis of the eigendisplacements are presented as well. 

 \begin{table*}[ht]

\begin{center}
\hspace*{-2cm} \begin{tabular}{c c r | c c r}
 \hline \hline
 \multicolumn{3}{c}{CDP} & \multicolumn{3}{c}{DCDP} \\
   \hline
    \multicolumn{2}{c}{A modes (TO-LO) [cm$^{-1}$]} & Assignment& \multicolumn{2}{c}{A modes (TO-LO) [cm$^{-1}$]} &  Assignment  \\
    \hline
 This work & Exp.$^a$ & & This work & Exp.$^a$ & \\
  \cline{1-2} \cline{4-5}

 47 & 46& Latt. & 47 & 45 & Latt.  \\
 58 & 54 & Latt.  & 58 & 54 & Latt.  \\
 64-66 & 74& Latt.   & 64-66 & 74 & Latt.  \\
 78 & 79& Latt.  & 77 & 79 & Latt.   \\
 82-85 & ?& Latt.  & 81-85 & ? &  Latt.  \\
 105-113 & ? & Latt.  & 104-112 & ? & Latt.  \\
 127-131 & 122-130 & Latt. & 126-131 & 122-122 &Latt. \\
 232-257 & 205-249 & Latt. & 226-251 & 215-238 & Latt. \\
 348 & 365-? & Rot. PO$_4$ & 342 & 360-362 & Rot. PO$_4$\\
 376-379 & 388-392 & Stretch. PO$_4$ & 372-375 & 385-388 & Stretch. PO$_4$ \\
 458-488 & 470-505& Stretch. PO$_4$ & 453-481 & 480-505 & Stretch. PO$_4$ \\
 525 & 543 & Stretch. PO$_4$ & 518 & 533 & Stretch. PO$_4$ \\
 547-548 & 563 & Stretch. PO$_4$ & 532-533 & 542 & Stretch. PO$_4$  \\
 885-888 & 910$^b$ & Stretch. PO$_4$ & 637-639 & 628$^b$& Bend. (H-O---H)$_{\text{long}}$ \\
 891-900 & 869-880  & Bend. (H-O---H)$_{\text{long}}$ & 727-737 & 702$^b$& Bend. (H-O---H)$_{\text{short}}$  \\
 923-951 & 921 &  Stretch. PO$_4$ & 872 & ?/881 & Mixed \\
 979-1010 & 993-1051 & Stretch. PO$_4$  & 879 & 881/? & Mixed \\
 1017-1038 & ?-981 & Bend. (H-O---H)$_{\text{short}}$ & 910-948 & 955-970 & Bend. (H-O---H)$_{\text{short}}$  \\ 
 1112-1121 &1134-1142 & Stretch. PO$_4$  & 938-982 & 918-933 & Stretch. PO$_4$ \\
 1209-1210 &1219-1221 & Bend. (H-O---H)$_{\text{long}}$ & 977-1024 & 1011-1061 & Stretch. PO$_4$  \\
 1278-1284 & 1256-1264 & Bend. (H-O---H)$_{\text{short}}$& 1118-1126 & 1155-1165 & Stretch. PO$_4$  \\
 2193-2227 & 1800 & Stretch. (H-O---H)$_{\text{short}}$ & 1636-1652 & 1970/1720 & Stretch. (H-O---H)$_{\text{short}}$  \\
 2731 & 2680 & Stretch. (H-O---H)$_{\text{long}}$ & 2009 & 2090 & Stretch. (H-O---H)$_{\text{long}}$ \\
 \hline \hline
 \multicolumn{4}{l}{$^a$ experimental Raman 80K data from Ref. \onlinecite{Marchon1983}.} \\
 \multicolumn{4}{l}{$^b$ experimental infrared reflectivity 80K data from Ref. \onlinecite{Marchon1983}.} \\

\end{tabular}
 \caption{Zone-center phonon modes of A symmetry computed with PBE-D3 for ferroelectric CDP and DCDP. The LO-TO splitting of A modes in only observed in the $b^*$ direction. Experimental Raman measurements
 are also reported \cite{Marchon1983}. Question marks refers to unavailable/undetected phonon modes. Two experimental values separated by a slash sign indicate an uncertainty on the experimental assignment. \label{PhononCDP}}
  \end{center}
\end{table*} 

\begin{table*}[ht]
\begin{center}
\begin{tabular}{c  c c r | c c  c r}
 \hline \hline
 \multicolumn{4}{c}{CDP} & \multicolumn{4}{c}{DCDP} \\
   \hline
    \multicolumn{3}{c}{B modes  [cm$^{-1}$]}  & Assignment& \multicolumn{3}{c}{B modes [cm$^{-1}$]} &  Assignment  \\
    \hline
 TO & LO($a^*$) & LO($c^*$)& &TO &LO($a^*$) & LO($c^*$) & \\
  \cline{1-3} \cline{5-7}
 50 & 51 & 51& Latt. & 50 & 50 &50  & Latt.    \\
 75 & 75  &75 &Latt. & 75 &75  &75& Latt. \\
 81 & 84  &82 &Latt. & 81 & 84 &82 & Latt. \\
 91 & 105  &105& Latt. & 91 &104 &104 & Latt.  \\
 113 & 116  &116& Latt. & 113 & 115 &116 & Latt.    \\
 143 & 143 &148 & Latt.& 141 & 142 &146 & Latt. \\
 224 & 230  &230 & Latt. & 221 & 228 &228 & Latt.   \\
 376 & 377 & 377& Stretch. PO$_4$&368 & 369 &369 & Stretch. PO$_4$ \\
 424 & 432  & 431 &Stretch. PO$_4$& 419 & 425 &425 & Stretch. PO$_4$ \\
 478 & 487  & 483 &Stretch. PO$_4$& 466 & 477 &472 & Stretch. PO$_4$ \\
 504 & 532 & 527 &Stretch. PO$_4$& 494 &519   &518 & Stretch. PO$_4$ \\
 543 & 544 & 543 &Stretch. PO$_4$& 533 & 537 &535 & Stretch. PO$_4$\\
 864 & 865  & 865 &Stretch. PO$_4$& 639 &639 &639 & Bend. (H-O---H)$_{\text{long}}$\\
 891 & 891  & 891 &Bend. (H-O---H)$_{\text{long}}$ & 754 &756 &758 & Bend. (H-O---H)$_{\text{short}}$  \\
 920 & 955  & 942 &Stretch. PO$_4$& 846 & 848 &847 & Stretch. PO$_4$\\
 1040 & 1044   &1040 &Stretch. PO$_4$& 873 &879 &876 & Bend. (H-O---H)$_{\text{long}}$\\
 1050 & 1051 &1052 &Bend. (H-O---H)$_{\text{short}}$& 911 & 942 &932 & Stretch. PO$_4$ \\
 1101 & 1112 &1137 &Stretch. PO$_4$ & 982 & 984 &982 & Bend. (H-O---H)$_{\text{short}}$    \\ 
 1201 &1215   &1216&Bend. (H-O---H)$_{\text{long}}$& 1041 & 1047 &1042 & Stretch. PO$_4$ \\
 1325 & 1326  &1327 &Bend. (H-O---H)$_{\text{short}}$& 1099 & 1123 & 1146 & Stretch. PO$_4$  \\
 2345 & 2476  &2448 & Stretch. (H-O---H)$_{\text{short}}$&1729 & 1819 & 1797 & Stretch. (H-O---H)$_{\text{short}}$ \\
 2725 & 2735  &2726 &Stretch. (H-O---H)$_{\text{long}}$&1997 & 2004 & 1998 & Stretch. (H-O---H)$_{\text{long}}$  \\
 
 \hline \hline

\end{tabular}
 \caption{Zone center phonon modes of B symmetry computed with PBE-D3 for ferroelectric CDP and DCDP.  \label{PhononCDP2}}
  \end{center}
\end{table*}

For the following, we define $a^*$, $b^*$ and $c^*$ as the reciprocal space primitive vectors. For the A modes, we only observe LO-TO splitting along the $b^*$ direction, in agreement with the observation of Marchon and Novak.
For the B modes, the coupling with the electric field is observed alongside both $a^*$ and $c^*$. The most important effect (131 cm$^{-1}$ for CDP and 90 cm$^{-1}$ for DCDP) is observed for
the stretching of (H-O---H)$_{\text{long}}$ [(D-O---D)$_{\text{long}}$ for DCDP] along the $a^*$ direction. In addition, the difference between the CDP and DCDP phonon frequencies 
is mostly centered on the stretching and bending modes involving the O-H (O-D) bonds. The ratio between CDP and DCDP corresponding phonon frequencies
is always smaller than $\sqrt{2}$, with the highest value reaching 1.36 for the (H-O---H)$_{\text{long}}$ stretching mode and 1.34 for the (H-O---H)$_{\text{short}}$ stretching mode. 

For DCDP, it was impossible to clearly assign the modes at 872 and 879 cm$^{-1}$, for which the eigenmodes correspond to a mixing of (D-O---D)$_{\text{long}}$ bending and PO$_4$-group stretching.

Overall, a good agreement is observed between our computation and the experimental data reported by Marchon and Novak, except for the stretching mode of
 (H-O---H)$_{\text{short}}$. Indeed, we obtain 2193 cm$^{-1}$ and 1636 cm$^{-1}$ for CDP and DCDP, respectively, to be compared to the experimental value of 1800 cm$^{-1}$ and 1970/1720 cm$^{-1}$. 
We suspect that this large discrepancy is due to a miss-assignment of the experimental Raman peak, which could also corresponds to a double-resonant peak combining the bendings of both 
(H-O---H)$_{\text{short}}$ and (H-O---H)$_{\text{long}}$. Our interpretation is further supported by the fact that the ratio between this reported phonon mode in CDP and DCDP is 
either equal to 0.91 (using 1970 cm$^{-1}$ for DCDP) or 1.04 (using 1720 cm$^{-1}$), really far from the expected value of $\sqrt{2}$.

Instead, we would rather assign the experimental Raman peak at 2250 cm$^{-1}$ to this (H-O---H)$_{\text{short}}$ stretching (assigned previously to a double-resonant Raman in Ref. \onlinecite{Marchon1983})
and similarly the Raman peak at 1720 cm$^{-1}$ to this stretching for DCDP.
 
 The remaining discrepancies can be attributed to anharmonic effects, to the approximated treatment of the vdW, to the exchange-correlation and to the experimental error bar
 ($\pm$3 cm$^{-1}$ according to Ref. \onlinecite{Marchon1983}).
%


 \subsection{Dielectric properties}

 In addition, we investigate the dielectric properties of CDP. PBE-D3 predicts the following optical dielectric constants
 $\epsilon^{\infty}_{xx}=2.45$, $\epsilon^{\infty}_{yy}=2.41$, $\epsilon^{\infty}_{zz}=2.51$ and $\epsilon^{\infty}_{xz}=-0.02$. To the best of our knowledge, no experimental value 
 has been reported for the ferroelectric phase of CDP. It is expected that our predictions overestimate the real optical dielectric constants due to the well-known underestimation of the electronic gap within
 DFT.
 The components of the static dielectric permittivity, which include the ion contributions, are $\epsilon^{0}_{xx}=6.11$,  $\epsilon^{0}_{yy}=6.57$,  $\epsilon^{0}_{zz}=5.60$ and $\epsilon^{0}_{xz}=-0.29$.
 
The total polarization of a system can be obtained
using the Berry phase formulation \cite{King-Smith1993,Vanderbilt1993,Souza2002,Nunes2001}. The spontaneous polarization of CDP is computed by taking the difference of total polarization between
the ferroelectric and paraelectric phases.
For CDP and DCDP, the calculated spontaneous polarization is 55.4 mC/m$^2$ along $b$ axis, in excellent agreement with both the computation of Lasave \textit{et al.} \cite{Lasave2016} (54 mC/m$^2$) 
and the experimental observation \cite{Deguchi1982}. The other components are negligible.


The analysis of Born effective charge tensors can give further insights on the ferroelectric properties of CsH$_2$PO$_4$ crystal. 
They are reported in Table \ref{BEC_CDP}. The ones of DCDP are equivalent in the harmonic approximation.

\begin{table}[ht]
\begin{center}
{\footnotesize 
\begin{tabular}{c | c c c | c | c c c}
\hline \hline
\multicolumn{8}{c}{Born effective charge [e]} \\
\hline 
 Cs(1) & $\partial/\partial R_{\kappa 1}$ & $\partial/\partial R_{\kappa 2}$ & $\partial/\partial R_{\kappa 3}$ &
 P(1) & $\partial/\partial R_{\kappa 1}$ & $\partial/\partial R_{\kappa 2}$ & $\partial/\partial R_{\kappa 3}$ \\
 \hline
 $\partial/\partial {\cal{E}}_1$ & 1.34 &0.0& 0.04 &  $\partial/\partial {\cal{E}}_1$ & 3.07 & -0.05 &-0.14  \\
 $\partial/\partial {\cal{E}}_2$ & 0.0 & 1.36 & 0.0 & $\partial/\partial {\cal{E}}_2$ & -0.03 & 3.35 & -0.02 \\
 $\partial/\partial {\cal{E}}_3$ & 0.04 & 0.0 & 1.36 &  $\partial/\partial {\cal{E}}_3$ & 0.27 & -0.02 & 3.32 \\
 \hline \hline
 O(1) & $\partial/\partial R_{\kappa 1}$ & $\partial/\partial R_{\kappa 2}$ & $\partial/\partial R_{\kappa 3}$ &
 O(2) & $\partial/\partial R_{\kappa 1}$ & $\partial/\partial R_{\kappa 2}$ & $\partial/\partial R_{\kappa 3}$ \\
 \hline
 $\partial/\partial {\cal{E}}_1$ & -1.71 &0.04 & 0.63 &  $\partial/\partial {\cal{E}}_1$ & -1.14 & -0.02 &-0.06  \\
 $\partial/\partial {\cal{E}}_2$ & 0.06 & -0.90 & 0.0 & $\partial/\partial {\cal{E}}_2$ & 0.01 & -1.18 & 0.04  \\
 $\partial/\partial {\cal{E}}_3$ & 0.58 & 0.0 & -1.95 &  $\partial/\partial {\cal{E}}_3$ & -0.04 & 0.01 & -2.64 \\
  \hline \hline
 O(3) & $\partial/\partial R_{\kappa 1}$ & $\partial/\partial R_{\kappa 2}$ & $\partial/\partial R_{\kappa 3}$ &
 O(4) & $\partial/\partial R_{\kappa 1}$ & $\partial/\partial R_{\kappa 2}$ & $\partial/\partial R_{\kappa 3}$ \\
 \hline
 $\partial/\partial {\cal{E}}_1$ & -1.86 &-0.77 & 0.00 &  $\partial/\partial {\cal{E}}_1$ & -1.86 & 0.67 &-0.02  \\
 $\partial/\partial {\cal{E}}_2$ & -0.60 & -1.79 & -0.17 & $\partial/\partial {\cal{E}}_2$ & 0.53 & -1.79 & 0.17  \\
 $\partial/\partial {\cal{E}}_3$ & 0.04 & -0.12 & -1.09 &  $\partial/\partial {\cal{E}}_3$ & -0.01 & 0.17 & -1.24 \\
  \hline \hline
 H(1) & $\partial/\partial R_{\kappa 1}$ & $\partial/\partial R_{\kappa 2}$ & $\partial/\partial R_{\kappa 3}$ &
 H(2) & $\partial/\partial R_{\kappa 1}$ & $\partial/\partial R_{\kappa 2}$ & $\partial/\partial R_{\kappa 3}$ \\
 \hline
 $\partial/\partial {\cal{E}}_1$ & 0.36 & 0.00 & -0.29 &  $\partial/\partial {\cal{E}}_1$ & 1.80 & 0.95 &-0.17  \\
 $\partial/\partial {\cal{E}}_2$ & 0.00 & 0.40 & -0.00 & $\partial/\partial {\cal{E}}_2$ & 0.46 & 0.54 & -0.14  \\
 $\partial/\partial {\cal{E}}_3$ & -0.47 & -0.00 & 1.86 &  $\partial/\partial {\cal{E}}_3$ & -0.41 & -0.33 & 0.38 \\
 \hline
\end{tabular}
}
\end{center}
\caption{Born effective charges of CsH$_2$PO$_4$ computed with PBE-D3 expressed in cartesian coordinates. 
Only the first 8 atoms of the crystal are reported in this table; the 8 other atoms can be recovered by symmetry. \label{BEC_CDP}}
\end{table}

Overall, the computed diagonal components of the Born effective charge tensors for Cs and P atoms are close to their corresponding nominal charge while those of O and H atoms
depend strongly on the direction. Indeed, the $xx$ component of the Born effective charge tensor for H$_2$ atom (aligned with the $a$ axis) is 80\% larger than its corresponding nominal value, similarly
to the $zz$ component for H$_1$, indicating a strong polarizability of the hydrogen bonds.

These Born effective charges differ strongly from those reported by Lasave \textit{et al.} \cite{Lasave2016} i.e. $Z_{yx}$(H$_{2}$)=1.6e, $Z_{yy}$(H$_{2}$)=0.8e and $Z_{yy}$(P$_{1}$)=2.3e.
However, these discrepancies could be explained by the difference of 
the long-range e$^-$-e$^-$ treatment (PBE-D2 vs PBE-D3), by the fact that their theoretical calculations were performed at experimental lattice parameters instead of their relaxed values, 
or by the pseudopotentials used. 


%
  
  \subsection{Elastic properties}
  
   The calculated elastic and piezoelectric tensors of ferroelectric CDP are reported in Table \ref{elastic}. The DCDP elastic and piezoelectric tensors are equivalent.
   To the best of our knowledge, no experimental measurements of the elastic constants nor the piezoelectric constants have been performed on the ferroelectric phase of CDP. However for the sake of comparison, we
   report the room temperature experimental measurements for the paraelectric $P2_1/m$ phase of CDP \cite{Prawer1985}. \\
   
   The eigenvalues of this elastic tensor are all positive, pointing out its mechanical stability. While the predicted diagonal elastic constants are somehow close to the corresponding room temperature values,
the off-diagonal elastic constants diverge more from these experimental measurements, especially $e_{25}$ and $e_{35}$. Most of the discrepancies may be explained by temperature effects close to room temperature,
which seem particularly important
for this material \cite{Luspin1997}, as well as by the phase transition that occurs at 154.5K. Concerning the predicted piezoelectric coefficients, they are comparable in magnitude than the ones of ZnO \cite{Hamann2005b},
and are approximatively one to two orders of magnitude smaller than the ones predicted for rhombohedral BaTiO$_3$ \cite{Hamann2005b}.

   
  

\begin{table}[h]
\begin{center}
\begin{tabular}{c c | c c c c | c c c c}
 \hline \hline
 \multicolumn{2}{c|}{Piezoelectric}& \multicolumn{8}{c}{\multirow{2}{*}{Elastic constants [GPa]}}\\
 \multicolumn{2}{c|}{constants [pC/N]}& \\
 \hline
 &&\multicolumn{4}{c|}{This work ($P2_1$)} & \multicolumn{4}{c}{Exp.\cite{Prawer1985} ($P2_1/m$)} \\
  $\tilde{d}_{12}$ & 5.32  & $e_{11}$ & 30.2 & $e_{23}$ & 13.3 & $e_{11}$ & 28.83& $e_{23}$ & 14.50 \\ 
  $\tilde{d}_{22}$ & -2.11  & $e_{22}$ & 36.2 & $e_{13}$ & 18.7 & $e_{22}$ & 26.67 & $e_{13}$ & 9.79 \\ 
  $\tilde{d}_{32}$ & -0.92  & $e_{33}$ & 85.7 & $e_{12}$ & 13.9 & $e_{33}$ & 65.45 & $e_{12}$ & 18.24 \\ 
  $\tilde{d}_{41}$ & 3.83   &$e_{44}$ & 9.00 & $e_{15}$ & 6.1 &$e_{44}$ & 8.10 & $e_{15}$ & 5.13 \\ 
  $\tilde{d}_{52}$ & -2.29 &$e_{55}$ & 14.5 & $e_{25}$ & 0.0 &$e_{55}$ & 5.20 & $e_{25}$ & 8.40 \\ 
  $\tilde{d}_{61}$ & 3.54  &$e_{66}$ & 12.9 & $e_{35}$ & -16.9 &$e_{66}$ & 9.17 & $e_{35}$ & 7.50 \\ 
  $\tilde{d}_{63}$  &1.21 && $e_{46}$ & -1.4 &&& $e_{46}$ & -2.25\\
 \hline  \hline
\end{tabular}
\caption{\label{elastic} Piezoelectric and elastic constants computed in PBE-D3 for $P2_1$ CDP alongside with room temperature experimental measurements \cite{Prawer1985}.}
\end{center}
\end{table}




\section*{Conclusion}

In this work, we have investigated in depth the structural, vibrational, mechanical and dielectric properties of CDP in both its ferroelectric and paraelectric phases. We find that the use of vdW corrections, included
in this work through Grimme's DFT-D methods, is important to describe properly the geometry of hydrogen-based ferroelectrics i.e. KH$_2$PO$_4$, RbH$_2$PO$_4$ and CsH$_2$PO$_4$.
Our stability studies point out the possible existence of a $P2_1$ (Z=4) phase at low temperature, 
yet unreported experimentally or theoretically. Phonon frequencies for the CDP ferroelectric phase are compared to their respective experimental ones. On average, an excellent agreement is obtained for both
the lattice and high-frequency modes which are dominated by hydrogen motions. We predict the B phonon frequencies for this phase as well as the dielectric, piezoelectric and elastic constants for this ferroelectric phase. This work sheds light
on the stability of CDP. We also report the implementation of the DFT-D contributions to elastic constants inside the \textsc{Abinit} software.

\section*{Aknowledgments}

The authors acknowledge technical help from J.-M.~Beuken and scientific discussions with Ya. Shchur. This work was supported by the FRS-FNRS through a FRIA
Grant (B.V.T.) and the Communauté française de Belgique through the BATTAB project (ARC 14/19-057). Computational resources have been provided by the supercomputing facilities of
the Université catholique de Louvain (CISM/UCL) and the Consortium des
Equipements de Calcul Intensif en Fédération Wallonie Bruxelles (CECI) funded by
the Fonds de la Recherche Scientifique de Belgique (FRS-FNRS) under convention 2.5020.11.
The present research benefited from computational resources made available on the Tier-1 supercomputer of the Fédération Wallonie-Bruxelles, 
infrastructure funded by the Walloon Region under the grant agreement n$^\circ$1117545.


\appendix

\section{DFT-D contribution to elastic constants} \label{Impl}

The DFT-D methods introduce a pair-wise correction $E_{disp}^{(2)}$, independent of the density, that is added to the DFT energy to mimic the vdW interactions. In contrast with the derivation
of the interatomic force constants -presented in a previous paper \cite{Vantroeye2016}, no dependency with respect to the cell index is involved in the derivation of elastic constants. In
consequence, the $\kappa$ index refers to the collection of all atoms which are replica from atom $i$ in the reference cell in contrast to the notations of Ref. \onlinecite{Vantroeye2016}.

For strain response properties, it is easier to work with the energy per undeformed unit cell volume $\Omega_0$, ${\cal{E}}_{vol}$ defined as
\begin{equation}
 {\cal{E}}_{vol}(R_{\kappa\mu},\epsilon_{\alpha\beta}) = \frac{E_{cell}}{\Omega_0} \label{Vol}
\end{equation}
with $R_{\kappa\mu}$ an atomic displacement and $\epsilon_{\alpha\beta}$ a strain. 
When an electric field is applied, the quantity which has to be minimized is no more this volumetric energy, but the electrical enthalpy \cite{Hamann2005b}. However, as the DFT-D methods do not add direct
contributions to electric response properties (the change is only indirect, through the change of lattice parameters or interatomic force constants), we will limit ourselves to the volumetric energy. The minimization
of Eq. (\ref{Vol}) gives then the ground-state geometry.

The denomination 'elastic tensor' ${e}_{\alpha\beta,\gamma\delta}$ computed in DFPT is usually used to refer to the full second derivative of the volumetric energy with respect to strains:
\begin{multline}
  e_{\alpha\beta,\gamma\delta} = \frac{\mathrm{d}^2 {\cal E}_{vol}}{\mathrm{d}\epsilon_{\alpha\beta}\mathrm{d}\epsilon_{\gamma\delta}} =  
 \overbrace{\frac{\partial^2 {\cal{E}}_{vol}}{\partial \epsilon_{\alpha\beta}\partial \epsilon_{\gamma\delta}}}^{\text{clamped-ion elastic tensor } \bar{e}_{\alpha\beta\gamma\delta}} \\
   \underbrace{+
 2\sum_{\kappa\mu}\frac{\partial^2 {\cal{E}}_{vol}}{\partial R_{\kappa\mu} \partial \epsilon_{\alpha\beta}} \frac{\partial R_{\kappa\mu}}{\partial \epsilon_{\gamma\delta}}+
 \sum_{\kappa\mu,\kappa'\nu} \frac{\partial^2 {\cal{E}}_{vol}}{\partial R_{\kappa\mu} \partial R_{\kappa'\nu}} \frac{\partial R_{\kappa\mu}}{\partial \epsilon_{\alpha\beta}}
 \frac{\partial R_{\kappa'\nu}}{\partial \epsilon_{\gamma\delta}}}_{\text{internal relaxation contribution } \tilde{e}_{\alpha\beta\gamma\delta} }.
 \\
\end{multline}

The internal-relaxation contributions to the elastic tensor can be expressed as \cite{Hamann2005b}
\begin{equation}
 \tilde{e}_{\alpha\beta,\gamma\delta} = - \frac{1}{\Omega_0} \sum_{\kappa\mu,\kappa'\nu} \Lambda_{\kappa\mu,\alpha\beta} \left(C^{-1}\right)_{\kappa\mu,\kappa'\nu}\Lambda_{\kappa'\nu,\gamma\delta},
\end{equation}
where $\left(C^{-1}\right)_{\kappa\mu,\kappa'\nu}$ is the pseudoinverse of $C_{\kappa\mu,\kappa'\nu}$ -the interatomic force constants in reciprocal space at zone center- and $\Lambda_{\kappa\mu,\alpha\beta}$ the internal strain coupling parameter, defined as
\begin{equation}
\Lambda_{\kappa\mu, \alpha \beta} = - \Omega_0 \frac{\partial^2 {\cal E}_{vol}}{\partial R_{\kappa\mu}\partial \epsilon_{\alpha \beta}}.
\end{equation}

The usual DFT derivation of the previously-introduced quantities (clamped-ion elastic tensor, internal strain coupling parameters, etc.) 
can be found elsewhere \cite{Hamann2005,Hamann2005b,Hamann2005c}. For the DFT-D contributions, one can show that a strain derivative of any pair-wise function $g(R_{AB})$ can be expressed as
\begin{equation}
 \frac{\partial} {\partial \epsilon_{\alpha\beta}} \left[ g(R_{AB}) \right]  =  \sum_{\kappa} R_{\kappa \beta}\frac{\partial} 
 {\partial R_{\kappa\alpha}} \left[ g(R_{AB}) \right].
\end{equation}
As a consequence, the DFT-D contribution to the clamped-ion elastic tensor is given by
 \begin{equation}
  \bar{e}^{disp}_{\alpha\beta,\gamma\delta} = \frac{1}{\Omega_0}\left[\sum_{\kappa} \sum_{\kappa'} \frac{\partial^2 E^{(2)}_{disp}}{\partial R_{\kappa\alpha}\partial R_{\kappa'\gamma}} R_{\kappa \beta} R_{\kappa' \delta} \right.
  \left. + \sum_{\kappa} \frac{\partial E^{(2)}_{disp}}{\partial R_{\kappa\alpha}} R_{\kappa\delta} \delta_{\beta \gamma} \right] \\
 \end{equation}
 and its contribution to the internal strain coupling parameters by
 \begin{equation}
  \Lambda^{disp}_{\kappa\mu, \alpha\beta} = -  \sum_{\kappa'} \frac{\partial^2 E^{(2)}_{disp}}{\partial R_{\kappa\mu}\partial R_{\kappa' \alpha}} R_{\kappa'\beta}.
 \end{equation}
 
Note however that this DFPT elastic tensor can differ from the ``real'' elastic tensor \cite{Hamann2005c}, defined as
\begin{equation}
 e^{real}_{\alpha\beta,\gamma\delta} = \frac{\partial \sigma_{\alpha\beta}}{\partial \epsilon_{\gamma\delta}} = \frac{1}{\Omega} \frac{\partial^2 E_{cell}}{\partial \epsilon_{\alpha\beta}\partial\epsilon_{\gamma\delta}}-
 \delta_{\gamma\delta} \sigma_{\alpha\beta},
\end{equation}
in the case in which the elastic tensor is computed away from the relaxed lattice parameters. In this last expression, $\Omega$ refers to the current volume of the primitive cell and 
$\sigma_{\alpha\beta}$ to the stress tensor.
 
 In DFT-D3, it is also possible to include a 3-body term (see for example Ref. \onlinecite{Grimme2010}). However, its contribution will be neglected in this work for the same reason
 as exposed in Ref. \onlinecite{Vantroeye2016}.
 
 
 The implementation of the elastic constants and internal strain coupling parameters 
 inside the \textsc{Abinit} software \cite{Abinit2005,Abinit2009,Gonze2016} completes the implementation of the DFT-D methods for first order response functions. 
 This implementation has been validated with respect to finite differences on distorted graphite: AB stacking, one atom moved by $10^{-5}$ in reduced coordinates,
 relaxed DFT-D3 in-plane and out-plane lattice parameters  i.e. 2.46~\AA $\,$ and 6.96 \AA,  the unique axis distorted by 5$^{\circ}$.
 The PBE functional was used for this validation. For the finite difference calculations, we used a first-order technique on the forces/stresses to get the clamped-ion and relaxed-ion tensors. 
Relative atomic displacements of $10^{-7}$ for the related derivatives and lattice change of $5\,10^{-5}\%$ for the strain derivatives were used.

Table \ref{TabComp} shows the comparison between finite differences and DFPT for the DFT-D3 contributions to the clamped-ion
elastic tensor and internal strain coupling parameters\footnote{using the DFPT expressions for the elastic tensors as defined previously in this work.}.
The agreement between finite difference and DFPT reaches up to 5 digits, thus validating validate the DFPT equations presented in this work. The implementation
of the internal relaxation contribution to the elastic constants and of the dispersion contribution to the interatomic force constants have already been discussed elsewhere \cite{Hamann2005b,Vantroeye2016}.

\begin{table}[h]
\begin{center}
\begin{tabular}{l c c}
\hline \hline
& Fin. diff. DFT-D & DFPT DFT-D \\
&  $\bar{e}_{\alpha\beta\gamma\delta}^{disp}$ [$\mu$Ha/Bohr$^3$] &  $\bar{e}^{disp}_{\alpha\beta\gamma\delta}$ [$\mu$Ha/Bohr$^3$] \\
 \\[-0.4cm]
\hline \\[-0.1cm]
$\bar{e}^{disp}_{33,33}$ & $\mathbf{-10.592197}49$ & $\mathbf{-10.592197}50$ \\                                              
$\bar{e}^{disp}_{12,33}$ &      $\mathbf{39.25711}2699$ & $\mathbf{39.25711}1270$ \\
$\bar{e}^{disp}_{12,12}$ &      $\mathbf{278.50914}204$ & $\mathbf{278.50914}508$ \\
 \\[-0.4cm]
 \hline \hline
& Fin. diff. DFT-D & DFPT DFT-D \\
&  $\Lambda_{\kappa\mu,\alpha\beta}^{disp}$ [$\mu$Ha/Bohr] &  $\Lambda_{\kappa\mu,\alpha\beta}^{disp}$ [$\mu$Ha/Bohr] \\
\hline
$\Lambda^{disp}_{13,33}$ & $\mathbf{10.05884}6720$ & $\mathbf{10.05884}2459$ \\                                              
$\Lambda^{disp}_{13,12}$ &      $\mathbf{-0.488955}908$ & $\mathbf{-0.488955}065$ \\
 \\[-0.4cm]
\hline \hline

\end{tabular}
\end{center}
\caption{Validation of our implementation by comparison of the dispersion contribution to clamped-ion elastic tensor
and internal strain coupling parameters (cartesian coordinates) in DFT-D3
computed by finite differences and by DFPT for distorded graphite. \label{TabComp}}
\end{table}

\end{document}